\documentclass[a4paper,aps,pra,twocolumn,preprintnumbers,amsmath,amssymb]{revtex4}

\usepackage{amsmath,amsfonts,amssymb,graphics,graphicx,epsfig,color,times}
\usepackage{graphicx}
\usepackage{epsfig}
\usepackage[english]{babel}

\newcommand{\be}{\begin{equation}}
\newcommand{\bea}{\begin{eqnarray}}
\newcommand{\eea}{\end{eqnarray}}
\newcommand{\ee}{\end{equation}}

\def\one{\ensuremath{\hbox{$\mathrm I$\kern-.6em$\mathrm 1$}}}

\def\qed{\leavevmode\unskip\penalty9999 \hbox{}\nobreak\hfill
     \quad\hbox{\leavevmode  \hbox to.77778em{%
               \hfil\vrule   \vbox to.675em%
               {\hrule width.6em\vfil\hrule}\vrule\hfil}}
     \par\vskip3pt}
\newcommand{\beaa}{\begin{eqnarray*}}
\newcommand{\eeaa}{\end{eqnarray*}}
\newcommand{\bma}{\begin{subequations}}
\newcommand{\ema}{\end{subequations}}
\newcommand{\Var}[1] {\textrm{Var}#1}

\def\one{{\bf 1}}

\def\noxrightarrow[#1]{\dodoublegroupempty\dodoxrightarrow{#1}}
\def\noxleftarrow [#1]{\dodoublegroupempty\dodoxleftarrow {#1}}
\def\dodoxrightarrow#1#2{\mathrel{{\domthxarr0359\rightarrowfill{#1}{#2}}}}
\def\dodoxleftarrow#1#2{\mathrel{{\domthxarr3095\leftarrowfill{#1}{#2}}}}

\begin{document}

\title{\bf Deterministic quantum teleportation between distant atomic objects}

\author{H. Krauter$^{1}$, D. Salart$^{1}$, C. A. Muschik$^{2}$, J. M. Petersen$^{1}$, H. Shen$^{1}$, T. Fernholz$^{3}$, and E. S. Polzik$^{1}$}

\affiliation{$^{1}$ Niels Bohr Institute, Copenhagen University,
Blegdamsvej 17, 2100 Copenhagen, Denmark,\\
$^{2}$ ICFO-Institut de Ci\`{e}ncies
Fot\`{o}niques, Mediterranean Technology Park, 08860 Castelldefels
(Barcelona), Spain,\\
$^{3}$ School of Physics \& Astronomy, The
University of Nottingham, Nottingham, NG7 2RD, UK}
\begin{abstract}
Quantum teleportation is a key ingredient of quantum
networks~\cite{bri98,riedmatten04} and a building  block for
quantum computation~\cite{gottesman99,gottesman01}. Teleportation
between distant material objects using light as the quantum
information carrier has been a particularly exciting goal. Here we propose and
demonstrate a new element of the quantum teleportation landscape,
the deterministic continuous variable teleportation between
distant material objects. The objects are macroscopic atomic
ensembles at room temperature. Entanglement required for
teleportation is distributed by light propagating from one
ensemble to the other. We demonstrate that the
experimental fidelity of the quantum teleportation is higher than
that achievable by any classical process. Furthermore, we
demonstrate the benefits of deterministic teleportation by
teleporting a sequence of spin states evolving in time from
one distant object onto another. The teleportation protocol is applicable to other important systems, such as mechanical oscillators coupled to light or cold spin ensembles coupled to microwaves.
\end{abstract}
\maketitle

Quantum teleportation of discrete~\cite{bennet93} and
continuous~\cite{vaidman94} variables is the transfer of a quantum
mechanical state without the transmission of a physical system
carrying this state. The first experimental teleportation
protocols employed light as the carrier of quantum
states~\cite{bouw1997,Furusawa98}. Teleportation of atomic states
over microns distances has been realized in two experiments using
short range interactions between trapped ions~\cite{riebe04}.
Interspecies teleportation from light onto atoms has been achieved
both deterministically for continuous variables~\cite{she06} and
probabilistically for discrete variables~\cite{Chen2008}. Recently,
probabilistic teleportation between two ions~\cite{Olmschenk2009}, atoms~\cite{Rempe2012} and atomic ensembles~\cite{pan12} over a macroscopic distance
has been demonstrated. While probabilistic teleportation, in which
entanglement is distributed by photon
counting~\cite{bouw1997,dua01}, is capable of reaching distances of
many km~\cite{riedmatten04,zeilinger12}, the power of continuous
variable (cv) teleportation is that it succeeds deterministically
in every attempt~\cite{Furusawa98,she06}, that it is capable of
teleporting complex quantum states~\cite{furusawa11}, and that it
can be used in universal quantum computation~\cite{gottesman01}.
Here, we propose and experimentally demonstrate for the first time the deterministic cv teleportation beetween two distant material objects thus extending the powerful cv
teleportation~\cite{Furusawa98,she06,furusawa11} onto atomic memory
states. The protocol which succeeds in every attempt allows us to
teleport dynamically changing quantum states of collective atomic
spins with the bandwidth of tens of Hz.

\begin{figure}
\includegraphics[width=1\columnwidth]{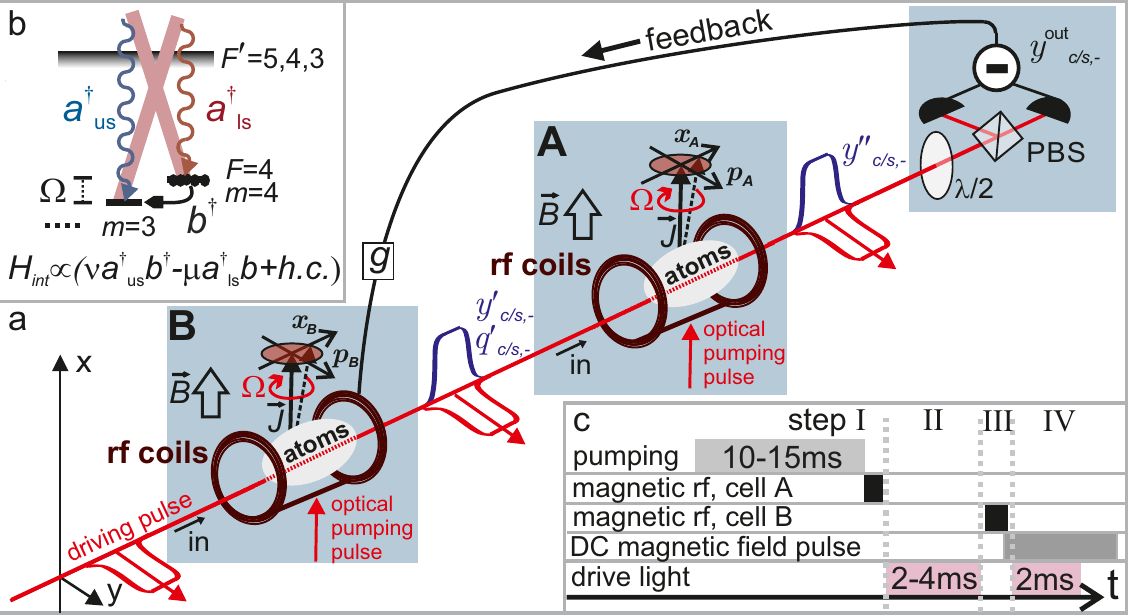}\caption{\textbf{Teleportation experiment}. (a) The experimental
layout.  A strong driving pulse propagates first through ensemble
B creating the modes $y',q'$ entangled with B and then through
ensemble A whose state is to be teleported. Joint measurements on
the modes $y''$ are performed by polarization homodyning.
Teleportation is completed by classical communication of these
results to B. More comments in the text. (b) The level scheme and
relevant transitions. Classical drive field (thick lines) and
quantum fields forming the modes $y,q$ (wavy lines) are shown, (c)
The time line of the experiment. I. Preparation of the input
state, II. Entanglement and joint measurement, III. Feedback, IV.
Read-out of the teleported state.\label{F1}}
\end{figure}

A quantum teleportation process begins with the creation of a
pair of entangled objects. In our experiment these two objects are
an atomic ensemble at site B and a photonic wave packet generated
by interaction of this ensemble with a driving light pulse
(Fig.~\ref{F1}a). The wave packet
travels to site A, the location of the atomic ensemble whose state
is to be teleported. This step establishes a quantum link between
the two locations. Following the interaction of the ensemble A and the wave packet, a
measurement is performed on the transmitted light. The results of this measurement are
communicated via a classical channel to site B, where they are fed
back via local operations on the second entangled object, i.e. the
ensemble B, thus completing the process of teleportation.
Cv teleportation is described in the language of
canonical operators $x,p$ for atoms and $y,q$ for light which obey the usual commutation relations $[x,p]=[y,q]=i$. A generic condition for a cv entangled state for Gaussian states~\cite{ham08} is $\text{var}(x-y)+\text{var}(p+q)< 2$. For atomic ensembles, fully spin polarized
along the x-axis, canonical operators are scaled dimensionless Cartesian
components of the collective spin: $x=J_{y}/\sqrt{|<J_x>|}$ and
$p=J_{z}/\sqrt{|<J_x>|}$~\cite{ham08}, where $J_{x,y,z}=\sum_{i}
j_{x,y,z}^i$ (summed over all atoms $i$) is the collective angular
momentum of the ensemble. Here, we employ $^{133}$Cs atoms initiated in a fully polarized $|F=4, m_F=4\rangle$ ground state. The usual link
between the ladder operator $b$ for collective atomic
excitations~\cite{ham08} of the state $m_F=3$ (Fig.~\ref{F1}b) and
canonical variables is $b^\dagger=(x-ip)/\sqrt{2}$.
Atoms are placed in a bias
magnetic field along the x-axis, so that in the lab frame the observables $x \propto J_y$ and
$p \propto J_z$ rotate at Larmor frequency $\Omega$ (Fig.~\ref{F1}a) according
to the atomic Hamiltonian
$H_{\text{Atomic}}=-\Omega\left({x}^2+{p}^2\right)/2$. Note that here we use the parallel orientation of the macroscopic spins of
the two ensembles (Fig. 1a) which is optimal for the teleportation protocol. It corresponds to the same sign of the Larmor frequency $\Omega$ in $H_{\text{Atomic}}$ for the two ensembles. This is to be compared to~\cite{ham08,kra10} were the antiparallel spin orintation, optimal for creating entanglement between two atomic spin ensembles, was used.

The atom-light
interaction is shown in Fig.~\ref{F1}b and involves two scattering
processes $H_{int} \propto \nu a_\textrm{us}^\dagger b^\dagger -
\mu a_\textrm{ls} ^\dagger b + h.c.$ where
$a^\dagger_\textrm{us/ls}$ generate photons in the upper/lower
($\omega_{0} \pm \Omega$) sideband modes of the driving field
$\omega_0$. The interaction $H_{int}$ contains both essential
ingredients of the teleportation protocol, the creation of
entanglement (the first term) and a
beam-splitter type operation between atoms and photons (the second
term)~\cite{ham08,kra10}. For our setting the ratio of the two terms is $\mu/\nu = 1.38$. The entanglement used in this protocol is between the atomic ensemble B and the light field sent to ensemble A.
The photons scattered forward into Larmor frequency sidebands populate the modes relevant for teleportation whose canonical variables $y_{c,s}$ and $q_{c,s}$  are $y_c \cos(\Omega t) + y_s \sin(\Omega t) \propto a_\textrm{us}e^{-i\Omega t}+a_\textrm{ls} e^{i\Omega t} +h.c.$ and similarly for $q$. The detailed theory of the protocol is presented in the Supplementary Information (SI), where exact definitions and properties of these modes are given in Eq.~(S3,S4). The generic form for them is $y_{c/s,f} \propto \int_0^T \cos/\sin (\Omega t)f(t) {y}(t)dt $, where $f(t)$ is a function which varies slowly on the time scale of the Larmor period.

The experiment (Fig.~\ref{F1}a) utilizes two room temperature gas
ensembles of Cesium
atoms in glass cells with spin protecting coating as
in~\cite{she07,ham08,was09} placed at a distance of $0.5$m.
Optical pumping initializes both ensembles
into the  $|F=4, m_F=4\rangle$ coherent
spin state (CSS )state with
$\textrm{Var}(J_y)\cdot\textrm{Var}(J_z)=J_x^2/4$ with $J_x\approx
4 N_A$ and $\langle J_y \rangle =0$ and $\langle J_z \rangle =0$,
corresponding to a vacuum state with variances
$\textrm{Var}(x)=\textrm{Var}(p)=1/2$. The spin of the ensemble A
to be teleported is then displaced with mean values $\langle
x_A\rangle$ and $\langle p_A\rangle$ by a weak radio-frequency
(rf) magnetic field pulse of frequency $\Omega$ corresponding to
the creation of coherent superposition of electronic ground states
$m_F=3,m_F=4$ (Fig.~\ref{F1}b).

The layout and the time sequence for teleportation and verification
are shown in Fig.~\ref{F1}a,c. A y-polarized, $3$ms long $5.6$mW light
pulse , blue
detuned from the $D_2$ line $F=4\rightarrow 5$ transition by
$\Delta=850$MHz drives the interaction. The forward scattered mode, x-polarized and
described by $y',q'$ is entangled with the collective spin B and
co-propagates with the drive light towards the site A. The interaction with the ensemble A leads to partial mapping of its state onto light and is followed by the Bell measurements on
the light modes  of the upper and lower sidebands performed via polarization homodyning with the driving light
acting as the local oscillator yielding $y_c''= (y_\textrm{us}''+y_\textrm{ls}'')/\sqrt 2$ and $y_s''=(q_\textrm{ls}''-q_\textrm{us}'')/\sqrt 2$ . The measurements of $y_c''$ and $y_s''$ serve as the joint measurement of ensemble A and the light coming from site B as can be seen directly from Eq. (S1) of the SI.   Near unity teleportation fidelity can be
achieved~\cite{sm}, if the driving fields for A and B ensembles are
made time-dependent. However, even with top hat driving pulses a
sufficiently high fidelity can be achieved, if an optimal temporal
mode for the detected homodyne signal is chosen. The optimal
readout mode is $y_{c/s,-} \propto \int_0^T \cos/\sin (\Omega t)
e^{-\gamma t} {y}(t)dt $, where
$T$ is the pulse duration and $\gamma$ is
the decay rate of the atomic state. Measurements of $y_{c/s,-}$
are conducted by electronic processing of the photocurrent. The
teleportation protocol is completed by sending the measurement
results $y^{out}_{c/s,-}$ via a classical link to the site B where
spin rotations in the y,z plane conditioned on these results are
performed using phase and amplitude controlled rf magnetic field
pulses at frequency $\Omega$.
The deterministic
character of the homodyne process ensures success of the
teleportation in every attempt.

\begin{figure}
\includegraphics[width=1\columnwidth]{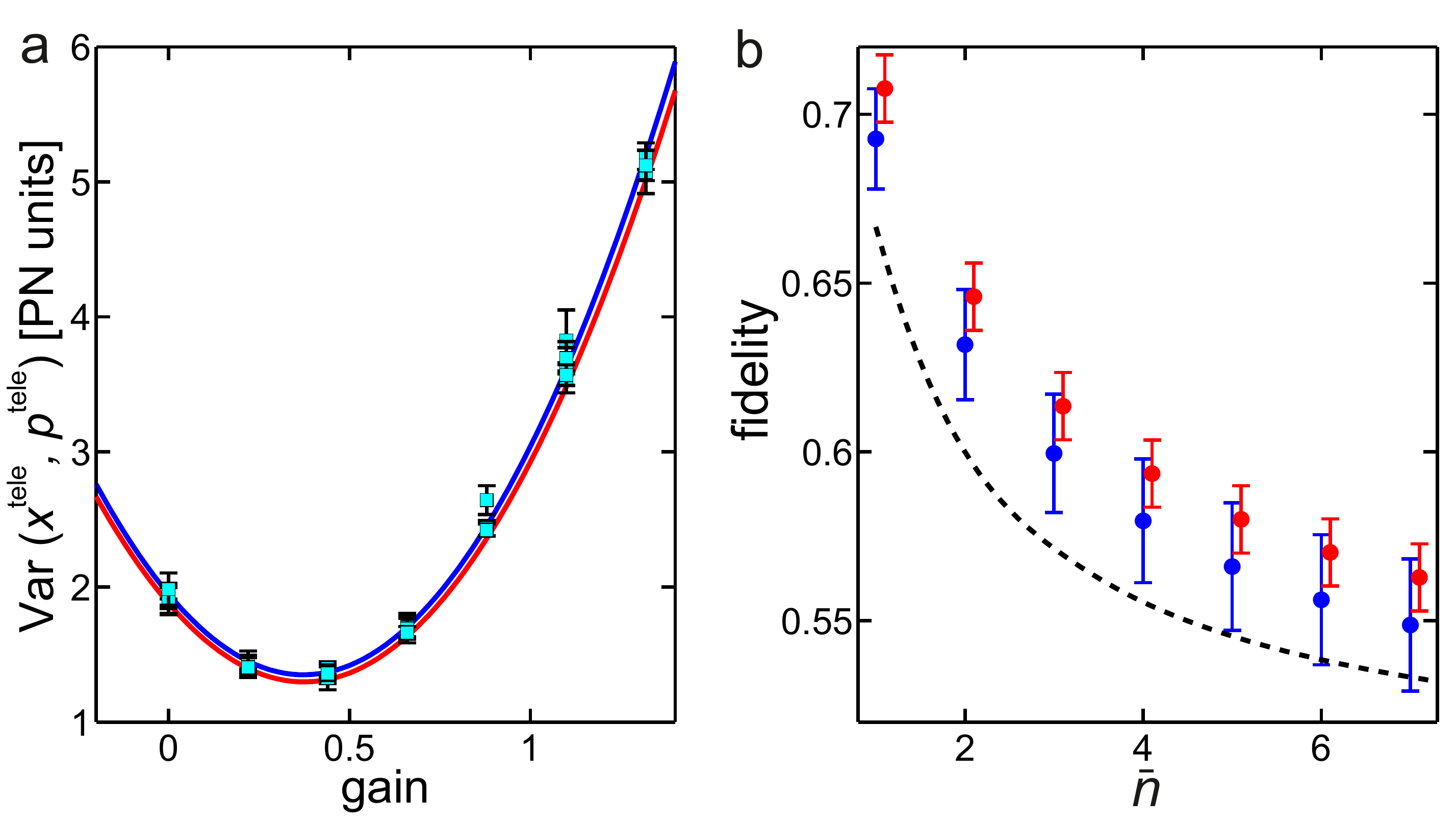}
\caption{\textbf{Teleportation fidelity}. (a) The variances of the
teleported state $\Var(x^\textrm{tele},p^\textrm{tele})$ in
projection (vacuum) noise (PN) units as a function of gain. Several data points for each gain correspond to various input states (the vacuum, the CSS with displacements of 5 in vacuum units and phases 0, $\pi/4$, $\pi/2$ and CSS with displacement 25 and phase 0). The error bars represent one standard deviation of the atomic variance for 5-10 subsets of 2000 points.
(b)Teleportation fidelity as a function of the mean photon number
of the Gaussian distribution of the input set of states. Blue curves/points - feedback by rf pulses applied to ensemble B, red curves/points - feedback
applied numerically to the read-out results of B (see comments in the text).  Black dashed line
represents the classical benchmark. For the error bars, the uncertainty of  kappa, shot noise, detection efficiency and the quadratic fit of the atomic variance vs gain were taken into account.\label{F2}}
\end{figure}
The quantum character of the teleportation is verified by
comparing the fidelity of state transfer to the classical
benchmark fidelity. More specifically, we perform
the teleportation using
various sets of coherent spin states of ensemble A with  varying $<J_{y,z}>$ , corresponding to displaced vacuum (coherent states)
in quantum optical terms, as input
states. For such states the individual state transfer
fidelity is calculated from the first two moments~\cite{sm}, i.e.
the mean values and the variances
$\sigma_x^2=\textrm{Var}(x_B^\textrm{tele})$,
$\sigma_p^2=\textrm{Var}(p_B^\textrm{tele})$. We then evaluate the average transfer fidelity for sets of coherent input states with a Gaussian distribution of displacements with mean number of spin excitations~\cite{ham05,sm} $\bar{n}=<b^\dagger b>$. A rigorous classical benchmark
fidelity $(1+\bar{n})/(1+2\bar{n})$ for transmission of such classes of states has
been derived in~\cite{ham05}. Demonstration of a fidelity above the
classical benchmark signifies the success of quantum teleportation
and is equivalent to the ability of the teleportation channel to
transfer entangled states.
For every input state, 10.000-20.000 teleportations have been
performed with one full
cycle of the protocol lasting $20$ms. Fig.~\ref{F2}a shows the
variance of the teleported states as a function of gain $g$. The quadratic dependence of the variances
on $g$ predicted by the model~\cite{sm} fits the experimental data
very well. For a certain range of
$g$ the atomic variances are reduced due to the
entanglement of the transmitted light with the ensemble
$B$~\cite{sm}. Fig.~\ref{F2}b presents the experimental fidelity
(blue dots), which is above the classical benchmark for $\bar{n}
\leq 7$.
The classical feedback conditioned on the Bell measurement result can be applied in two ways. It can be done by performing a displacement operation with an rf pulse applied to ensemble B, followed by a subsequent verification by the read-out of the atomic state. Alternatively, the verification read-out can be performed first, followed by the displacement operation applied to the result of the measurement numerically
(see SI).
In theory, those two procedures are
equivalent, but in the experiment the resulting fidelity for the
latter one is slightly higher (red dots in Fig.~\ref{F2}b) since
the application of rf fields required in the former procedure
introduces additional technical errors.

\begin{figure}
\includegraphics[width=1\columnwidth]{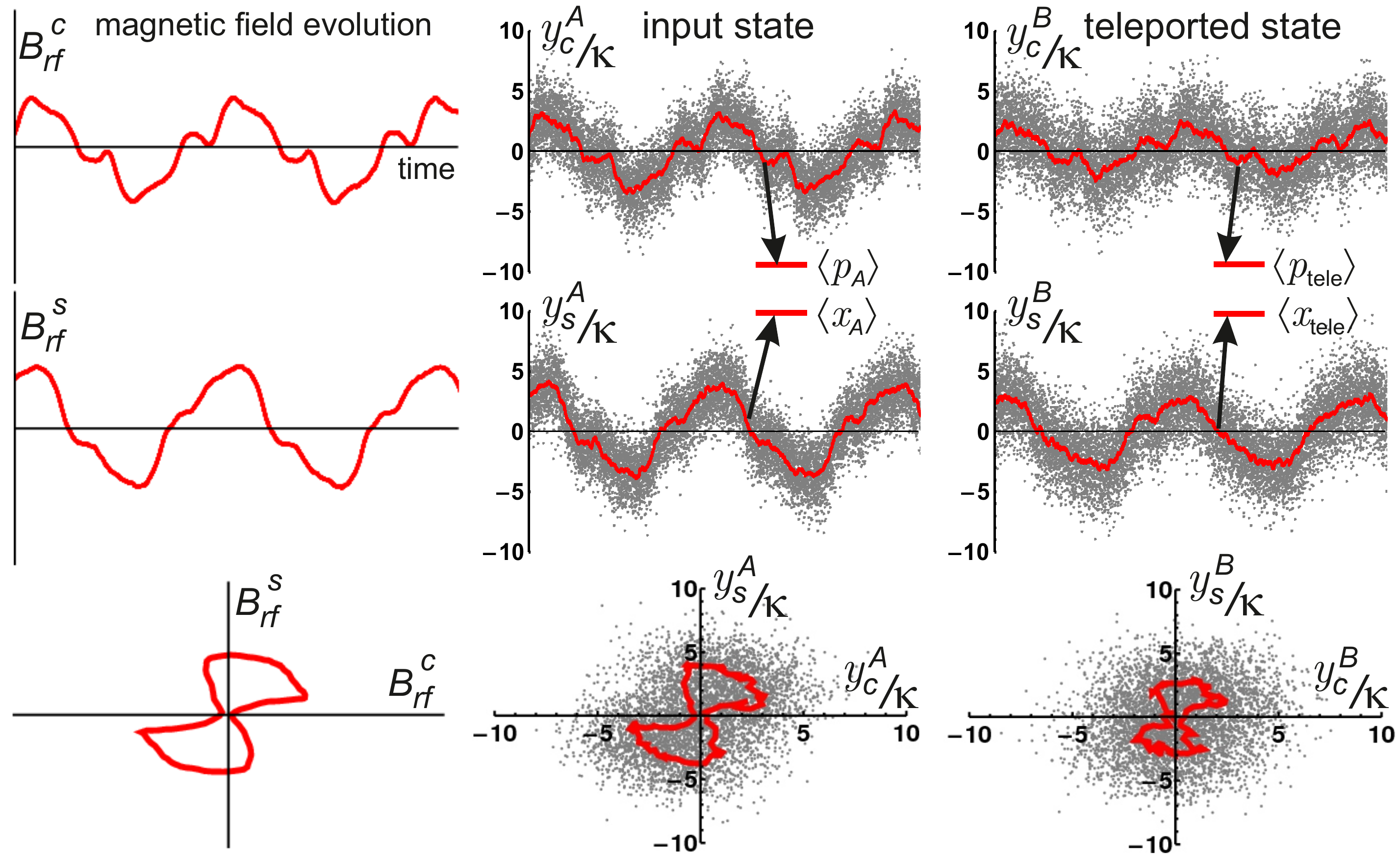}
\caption{\textbf{Teleportation of a sequence of spin states.} Left column - rf magnetic field applied to spin A with
components $x_A\propto J_y\propto B_{rf}^s, p_A\propto J_z\propto
B_{rf}^c$ with the amplitude of $B_{rf}\approx 1$ picoTesla. Center/right columns - the read-out of the
input/teleported spin states A/B in vacuum units.
Every point is one teleportation run with the points taken at
the rate of $\approx 50$Hz with the whole shown sequence taking $\approx 200$ sec. The lines present the running average
of the points. The first/second row is the $p_{A,B}/x_{A,B}$
variable and the third row is a two-dimensional plot
$x_{A,B},p_{A,B}$. The optimal teleportation gain for this
evolution is 0.8 which is seen as a smaller mean amplitude of the
teleported evolution compared to the original.\label{F3}}
\end{figure}

The deterministic teleportation can  be used for "stroboscopic"
teleportation of a sequence of spin states changing at a rate of $\approx 50$ Hz from A to B. To illustrate this attractive feature, we have
performed repeated teleportation cycles while varying the
amplitude and phase of the input state. The results are presented
in Fig.~\ref{F3}. The left column displays the time varying  rf field in the picoTesla range which is
applied to prepare a new spin state A in each individual teleportation run, after
initializing both ensembles to vacuum between the runs.
The central column shows the read-out of the input state evolution
of ensemble A and the right column shows the read-out of the
teleported state evolution. The points represent results for
individual teleportation runs.

The fidelity of the teleportation can be further improved by using
time varying drive pulses~\cite{sm} and increasing the optical
depth of the atomic ensembles. Cv teleportation is capable of
teleporting highly non-classical states as shown for teleportation
of light modes~\cite{furusawa11}, so it can be expected that
deterministic teleportation of an atomic qubit~\cite{dua01} can be
performed by developing the present approach. The stroboscopic
teleportation of spin dynamics can also be extended towards a true
continuous in time teleportation paving the way to teleportation
of quantum dynamics and simulations of the interaction between two
distant objects which have never interacted directly~\cite{muschik13}.
Cv atomic teleportation allows for performing quantum sensing at a remote location,
spatially separated from the location of the object. This teleportation protocol is, in principle, applicable to other systems described by strongly coupled harmonic oscillators, for example, to mechanical oscillators in a quantum regime coupled to light or cold spin ensembles coupled to microwaves.

\subsection{Methods Summary}
The Larmor precession of the atomic spin oscillators in the bias magnetic field allows us to perform quantum teleportation with a very large atomic object consisting of $N_A \approx 10^{11}-10^{12}$ atoms and to use a strong drive with the number of photons of $N_{ph} \approx 10^{13}-10^{14}$. The relative size of vacuum state fluctuations in a multiparticle ensemble scales as $N^{-1/2}$.  Therefore all technical fluctuations of spins and light must be reduced to $<<10^{-6}$ before the vacuum state noise level which is the benchmark for cv quantum information processing can be reached. We achieve this by encoding quantum states of atoms and light at the high Larmor frequency $\Omega=322$kHz (the bias magnetic field of $B\approx 0.9$G) where technical noise is much lower than at lower frequencies. This allows us to achieve vacuum (projection) noise level for atoms and vacuum (shot) noise level for light.
Using the strong driving field also as the local oscillator field for polarization homodyne detection of photonic variables $y_{c,s}$ allows us to use detectors with nearly unity quantum efficiency.

The calibration of the input atomic spin state, the joint
measurement, and the detection of the state teleported onto the
spin B are performed via polarization homodyning measurements of
the Stokes operator $S_2=(n_{+45}-n_{-45})/2$ given by the
difference of photon numbers polarized in $\pm 45^\circ$
directions (Fig.\ref{F1})~\cite{she07,muschik2012}. The measured
canonical variable for light is then defined as
$S_2\approx\sqrt{\Phi}/2\cdot y$ where $\Phi$ is the driving field
photon flux, which experimentally means that all measurements are
normalized to shot noise of light. The photocurrent is analyzed
with a lock-in amplifier at $\Omega$ and
further computer processed to obtain measurements of the temporal
modes of interest $y_{c/s,-}$. Light pulses for teleportation
and read-out always pass through both vapour cells
(Fig.~\ref{F1}). For the read out of each individual
ensemble the other ensemble is detuned from the atom-light
interaction by briefly detuning the $B$ field in the respective
cell. For off-resonant light well below saturation used here, the
linear transformation of light variables after dispersive
interaction with atoms is given by~\cite{kra10,muschik2012}:
\begin{equation}
y_{c/s,-}=\kappa \cdot p/x+c_{y}\cdot y_{c/s,f_y
}^\textrm{in}+c_{q}\cdot q_{s/c,f_q}^\textrm{in}+c_{N}\cdot
F_{p,x}. \label{recons}
\end{equation}
Here, the first term is a contribution of the atomic spin variable
due to Faraday rotation of light polarization, the second term is
proportional to the input value of the light quadrature $y$ of the
temporal mode $f_y$ and the third term is the contribution of the
other quadrature of input light $q$ of temporal mode $f_q$
resulting from back action of light on atoms~\cite{sm}. All input
light modes are always in a coherent or vacuum state with
$Var(y_{c/s,f_y }^\textrm{in})=Var(q_{c/s,f_q }^\textrm{in})=1/2$.
The last term in Eq.~\ref{recons} describes additional noise
arising from atomic decoherence with $Var(F_{p,x})=m/2$ with
$m=1.3$ found from the atomic spin relaxation~\cite{Vasilyev12}.
The value of the interaction constant $\kappa$ is found by
calibrating the Faraday rotation caused by the
ensemble~\cite{kra10}. The constants $c_y$ and $c_q$ are determined
by sending light with displacements of $< y_{c/s,f_y
}^\textrm{in}>$ and $<q_{c/s,f_q }^\textrm{in}>$, storing it in
the atomic medium, then reading it out onto another pulse
$<y_{c/s,-}^\textrm{out}>$ and measuring the ratios. The values
for $c_y$, $c_q$ and $c_N$ can also be calculated from the
model~\cite{sm}
based on three experimental parameters: the total transverse decay
rate of the atomic spin state $\gamma$, the contribution of spin
decoherence (spontaneous emission, collisions and inhomogeneity of
the magnetic field) to this decay rate $\gamma_\textrm{extra}$ and
$Z^{2} =(\mu+\nu)/(\mu-\nu)$. $Z^{2}=6.3$ is calculated from
Clebsch-Gordon coefficients for the atomic transitions and
experimentally verified~\cite{was09}. For $5.6$mW read out pulses of
$2$ms duration and room temperature Cs vapor pressure (effective
resonant optical depth of 34 for $22$mm long cells)
$\gamma=99.3\pm0.2$sec$^{-1}$ and
$\gamma_\textrm{extra}=26.3\pm0.2$sec$^{-1}$ have been measured. A
unitary contribution to the decay $\gamma - \gamma_\textrm{extra}$
is due to the collective coupling $H_\textrm{int}$, which
describes the rate of entanglement generation and the beam
splitter interaction~\cite{sm} and depends on the optical depth of
the ensemble, the optical detuning, and the intensity of the
driving field. The measured values of $\kappa, c_y, c_q$ agree
very well with the predictions of the model~\cite{sm} and are
$\kappa=0.87$, $c_y=0.93$, $c_q=0.50$ for our teleportation
setting. The last coefficient in the read-out equation can be
found from the measured parameters as $c_N=c_q \cdot \sqrt{2 \cdot
\gamma_\textrm{extra}/(\gamma -\gamma_\textrm{extra})}/Z=0.17$.
For the atomic state reconstruction the detection efficiencies
including optical losses $\eta_B=0.80\pm0.03$ and
$\eta_A=0.89\pm0.03$ for ensembles A/B are taken into account.

Using Eq.\ref{recons} the mean values of the input states of ensemble A are
found from measurements of light variables as $<x_A>=<y_{c,-}^{A}>/\kappa$,
$<p_A>=<y_{s,-}^{A}>/\kappa$ and their variances as
$\Var(x_A)=(\Var(y_{c,-}^{A})-c_y^2/2-c_q^2/2-c_N^2m/2)/\kappa^2$,
$\Var(p_A)=(\Var(y_{s,-}^{A})-c_y^2/2-c_q^2/2-c_N^2m/2)/\kappa^2$.

The prepared atomic input states are
found to be very close to ideal CSS with
$\Var(x_A)=\Var(p_A)=(1.03 \pm 0.03)\cdot 1/2$, which confirms the
validity of the read-out procedure. After each teleportation sequence, the
mean values and the variances
$<x_B^{\textrm{tele}}>,<p_B^{\textrm{tele}}>,\Var(x_B^{\textrm{tele}}),\Var(p_B^{\textrm{tele}})$
of the spin state of the target ensemble B are found in the same
way from the read-out of the verification pulse $y_{c/s,-}^{B}$.  For CSS input states with displacements of $0, 5,
25, 160$ in canonical units and phases $0,\pi/4, \pi/2$ in $x, p$
space, the variance of the teleported state showed no dependence on the displacement.
The experimental fidelity is determined
using a standard method of calculation of the state overlap~\cite{sm}.
Optimization of the teleportation protocol has been performed by
varying the drive pulse duration $T$, the measured temporal mode
of light, and the gain for the classical feedback.
The optimal read-out mode was always found with an exponential
decay rate equal to the spin decay $\gamma$ as expected from the
model.

\section{Acknowledgements}
We gratefully acknowledge discussions with J. I. Cirac, K.
Hammerer and D. V. Vasilyev. This work was supported by the ERC
grants INTERFACE and QUAGATUA, the Danish National Science
Foundation Center QUANTOP, the DARPA program QUASAR, the Alexander
von Humboldt Foundation, TOQATA (FIS2008-00784) and the EU
projects QESSENCE, MALICIA and AQUTE.

\section{Author Contributions}
H.K., D.S., J.M.P, H.S. and T.F. performed the experiment. The
theoretical model has been developed by C.A.M.\\
H.K, C.A.M., D.S. and E.S.P have written the paper. E.S.P. supervised the project.

\appendix
\renewcommand{\figurename}{Supplementary figure}
\setcounter{figure}{0} \setcounter{equation}{0}
\renewcommand{\thefigure}{S.\arabic{figure}}
\renewcommand{\theequation}{S.\arabic{equation}}
\widetext
\section*{Supplemental Information}

In the following, we provide additional information on the the read-out of the atomic spin state and the protocol used for teleportation. In Sec.~\ref{Sec:SI:Interaction}, we discuss the input-output relations for the interaction of a single cell with light. These equations are the basis for the read-out which is used for verification. In Sec.~\ref{Sec:SI:TeleportationScheme}, we explain the teleportation scheme, state the corresponding input-output relations for atoms and light and calculate the attainable fidelity.
\subsection{Light-matter interaction}\label{Sec:SI:Interaction}
We are interested in reading out the spin state of an atomic ensemble, which is described in terms of bosonic operators $x$ and $p$. This information is mapped to a coherent light field. The interaction of light with atoms, which are rotating in a magnetic field with a Larmor frequency $\Omega$ leads to temporally modulated light modes. More specifically, the atomic quadratures $x$ and $p$ are mapped to $\sin(\Omega t)$ and $\cos(\Omega t)$ modulated light modes respectively, which can be accessed individually. As explained below, we consider here specific light modes with an exponentially falling slowly varying envelope on top of the fast sine/cosine modulation. The relevant input-output relation for these read-out modes are given by
\begin{eqnarray}\label{Eq:SI:Read-out}
\left(
  \begin{array}{c}
   \!\! y_{\text{c},-} \!\!\\
   \!\! y_{\text{s},-} \!\!\\
  \end{array}
\right)\!\!&=&\!\!\kappa\left(
                                           \begin{array}{c}
                                            \!\! p^{\text{in}} \!\!\\
                                            \!\! x^{\text{in}} \!\!\\
                                           \end{array}
                                         \right)
\!+\!c_{\text{y}} \left(
  \begin{array}{c}
    y_{\text{c},f_{\text{y}}}^{\text{in}} \\
    y_{\text{s},f_{\text{y}}}^{\text{in}} \\
  \end{array}
\right)
\!+\!c_{\text{q}} \left(
  \begin{array}{c}
    -q_{\text{s},f_{\text{q}}}^{\text{in}} \\
    q_{\text{c},f_{\text{q}}}^{\text{in}} \\
  \end{array}
\right)
\!+\! c_{\text{N}} \left(
      \begin{array}{c}
        \!\! F_{p}^{\text{in}}\!\!\\
        \!\! F_{x}^{\text{in}}\!\!\\
      \end{array}
    \right)\!.
\end{eqnarray}
These equations include atomic decay. The first term on the right side is the desired atomic signal. The second and third term on the right represent contributions of the light field. The first subscript of the photonic operators refers to the fast modulation (i.e. modulation with $\sin(\Omega t)$ or $\cos(\Omega t)$), while the second subscript refers to the mode function of the slowly varying envelope (as explained below). $F_{x}^{\text{in}}$ and $F_{p}^{\text{in}}$ are atomic noise operators, $\kappa$, $c_{\text{y}}$, $c_{\text{q}}$ and $c_{\text{N}}$ are real coefficients and are given by Eqs.~(\ref{Eq:SI:coefficients}) below.\\
\\The interaction between atoms and light is governed by the Hamiltonian $ H=H_{\text{A}}+H_{\text{L}}+H_{\text{int}}$, where $H_{\text{L}}$ represents the free propagation of the light field along $z$ and $H_{\text{A}}$ is the free Hamiltonian of the atomic system. More specifically,
$
H_{\text{A}}=-\Omega\left(x^2+p^2\right)/2,
$
describes the Larmor precession with Larmor frequency $\Omega$ of the atoms in the magnetic field. The interaction Hamiltonian is a combination of a passive (beamsplitter-) part $H_{\text{BS}}$, which preserves the number of excitations in the system and an active (two mode squeezing) part $H_{\text{TMS}}$, which creates entanglement between atoms and light,
\begin{eqnarray}\label{Eq:Hamiltonian}
H_{\text{int}}&=&\sqrt{2\gamma_{\text{s}}}\left(\mu H_{\text{BS}}-\nu H_{\text{TMS}}\right)
=\sqrt{2\gamma_{\text{s}}}\left( Z p q(0)+\frac{1}{Z}x y(0)\right),
\end{eqnarray}
where $\mu=\frac{1}{2}\left(Z+\frac{1}{Z}\right)$, $\nu=\frac{1}{2}\left(Z-\frac{1}{Z}\right)$. Here, we assumed a pointlike atomic ensemble which is located at the origin $z=0$. The light field is described in terms of spatially localized modes~\cite{Silberfarb2003,Madsen2004,HPC05},
\begin{eqnarray}\label{Eq:LOcalizedLightModes}
y(z)&=&\frac{1}{4\pi}\int_b d\omega\left(a(\omega)e^{-i(\omega_0-\omega)z/c}+H.C.\right),\\
q(z)&=&\frac{-i}{4\pi}\int_b d\omega\left(a(\omega)e^{-i(\omega_0-\omega)z/c}-H.C.\right),\nonumber
\end{eqnarray}
where $c$ is the speed of light and $b$ is the bandwidth of the applied laser field with central frequency $\omega_0$. The canonical variables $y(z)$ and $q(z)$ obey the commutation relation $[y(z),q(z')]=c\delta(z-z')$. The width of the delta function is on the order of $c/b$. In the following, we write the time argument explicitly ($y(z,t)$, $q(z,t)$) and perform a variable transformation $\bar{y}(\xi,t)=y(ct-\xi,t)$ on the spatial argument of the light-field operators. The transformed operators describe the light field in the moving frame (see for example~\cite{HPC05,MKJPCP12}). Using this description, integrated light modes can be defined by considering an integral over the individual pieces of the light pulse with temporally varying
weighting functions (see Eq.~(\ref{Eq:SI:ExpModeNoisy}) and Eq.~(\ref{Eq:SI:GeneralModeDefinitions})).\\
\\We derive the input-output relations for the light-matter interaction given by  Eq.~(\ref{Eq:Hamiltonian}) in the limit $\Omega T \gg 1$, where $T$ is the total interaction time. Contributions which are on the order of $(\Omega T)^{-1}$ are neglected, which is a very good approximation for the experimental parameters considered here (compare~\cite{HPC05,HSP10,MPC11}). We include undesired noise processes, which lead to a decay of the transverse atomic spin at a rate $\gamma_{\text{extra}}$. In the presence of noise, the optimal slowly varying envelope for the read-out is an exponentially falling mode $e^{-\gamma t}$, where $\gamma=\gamma_{\text{s}}+\gamma_{\text{extra}}$.
In the experiment, the exponentially falling sine and cosine modulated light modes
\begin{eqnarray}\label{Eq:SI:ExpModeNoisy}
\left(
  \begin{array}{c}
    y_{\text{c},-} \\
    y_{\text{s},-} \\
  \end{array}
\right)&=&\frac{2\sqrt{\gamma}}{\sqrt{1-e^{-2\gamma T}}}\int_0^Tdte^{-\gamma t}\left(
                                                                                                                         \begin{array}{c}
                                                                                                                           \cos(\Omega t) \\
                                                                                                                           \sin(\Omega t) \\
                                                                                                                         \end{array}
                                                                                                                       \right)\bar{y}(ct,T)
\end{eqnarray}
are measured. The corresponding input-output relations are given by Eq.~(\ref{Eq:SI:Read-out}) above. The light and noise modes appearing in this equation are defined by
\begin{eqnarray}\label{Eq:SI:GeneralModeDefinitions}
\left(
  \begin{array}{c}
    y_{\text{c},f_{\text{y}}}^{\text{in}} \\
    y_{\text{s},f_{\text{y}}}^{\text{in}} \\
  \end{array}
\right)&=&\frac{1}{\sqrt{N_{\text{y}}}}\int_0^Tdt f_{\text{y}}(t)\left(
                                                              \begin{array}{c}
                                                                \cos(\Omega t) \\
                                                                \sin(\Omega t) \\
                                                              \end{array}
                                                            \right)\bar{y}(ct,0),\\
\left(
  \begin{array}{c}
    q_{\text{s},f_{\text{q}}}^{\text{in}} \\
    q_{\text{c},f_{\text{q}}}^{\text{in}} \\
  \end{array}
\right)&=&\frac{1}{\sqrt{N_{\text{q}}}}\int_0^Tdt f_{\text{q}}(t)\left(
                                                              \begin{array}{c}
                                                                \sin(\Omega t) \\
                                                                \cos(\Omega t) \\
                                                              \end{array}
                                                            \right)\bar{q}(ct,0),\nonumber\\
\left(
      \begin{array}{c}
        \!\! F_{p}^{\text{in}}\!\!\\
        \!\! F_{x}^{\text{in}}\!\!\\
      \end{array}
    \right)&=&\frac{1}{\sqrt{N_{\text{N}}}}\int_0^Tdt f_{\text{N}}(t)\left(
                                                              \begin{array}{c}
                                                                f_{p}(t) \\
                                                                f_x(t) \\
                                                              \end{array}
                                                            \right),\nonumber
\end{eqnarray}
where $f_x(t)$ and $f_p(t)$ are atomic noise operators with $\langle f_x(t)\rangle=\langle f_p(t)\rangle=0$ and $\langle f_x(t)f_x(t')\rangle=\langle f_p(t)f_p(t')\rangle=\frac{m}{2}\delta(t-t')$. Accordingly, $\text{var}\left(F_{x}^{\text{in}}\right)=\text{var}\left(F_{p}^{\text{in}}\right)=m/2$. For the interaction Hamiltonian, given by Eq.~(\ref{Eq:Hamiltonian}), one obtains $m=1.3$~\cite{NoisePaper,Denis}. The modulating functions $f_{\text{y}}$, $f_{\text{q}}$ and $f_{\text{N}}$ read
\begin{eqnarray*}
f_{\text{y}}(t)&=&\frac{1}{\sqrt{1-e^{-2\gamma T}}}\left(\left[2\sqrt{\gamma}\!-\!\frac{\gamma_{\text{s}}}{\sqrt{\gamma}}\right]e^{-\gamma t}\!+\!\frac{\gamma_{\text{s}}}{\sqrt{\gamma}}e^{-2\gamma T}e^{\gamma t}\right),\\
f_{\text{q}}(t)&=&Z^2\frac{\gamma_{\text{s}}}{\sqrt{\gamma}} \frac{1}{\sqrt{1-e^{-2\gamma T}}} \left(e^{-\gamma t}-e^{-2\gamma T}e^{\gamma t}\right),\\
f_{\text{N}}(t)&=&Z\frac{\sqrt{\gamma_{\text{s}}\gamma_{\text{extra}}}}{\sqrt{\gamma}}\frac{1}{\sqrt{1-e^{-2\gamma T}}}\left(e^{-\gamma t}-e^{-2\gamma T}e^{\gamma t}\right),
\end{eqnarray*}
with
\begin{eqnarray*}
N_{\text{y}}&=&\frac{1}{2}\int_0^Tdtf_{\text{y}}(t)^2,\ \ \ \
N_{\text{q}}=\frac{1}{2}\int_0^Tdtf_{\text{q}}(t)^2,\ \ \ \
N_{\text{N}}=\int_0^Tdtf_{\text{N}}(t)^2.
\end{eqnarray*}
The coefficients appearing in Eq.~(\ref{Eq:SI:Read-out}) are given by
\begin{eqnarray}\label{Eq:SI:coefficients}
\kappa=\frac{Z \sqrt{\gamma_{\text{s}}}}{\sqrt{2\gamma}}\sqrt{1-e^{-2\gamma T}},\ \ \ \
c_{\text{y}}=\sqrt{N_{\text{y}}},\ \ \ \ c_{\text{q}}=\sqrt{N_{\text{q}}},\ \ \ \ c_{\text{N}}=\sqrt{N_{\text{N}}}.
\end{eqnarray}
It is instructive to consider the limiting case of large $Z^2>>1$, with $\gamma_{\text{s}}T Z^2=\text{const.}$ In this limit, $\mu=\nu$ and we obtain a quantum non demolition (QND) interaction with $H_{\text{int}}\propto pq $. In the absence of decay, the coefficients of the input-output relation Eq.~(\ref{Eq:SI:Read-out}) become
\begin{eqnarray}\label{Eq:SI:coefficientsQND}
\kappa=Z \sqrt{\gamma T},\ \ \ \
c_{\text{y}}=1,\ \ \ \ c_{\text{q}}=\kappa^{2}/\sqrt{3},\ \ \ \ c_{\text{N}}=0.
\end{eqnarray}
In this case, the readout equations take a form extensively used previously~\cite{HPC05,HSP10}
\begin{eqnarray}\label{Eq:SI:readoutQND}
\left(
  \begin{array}{c}
    y_{\text{c},-} \\
    y_{\text{s},-} \\
  \end{array}
\right)
=\left(
  \begin{array}{c}
    y_{\text{c},-}^{\text{in}} \\
    y_{\text{s},-}^{\text{in}} \\
  \end{array}
\right)+\kappa\left(
                \begin{array}{c}
                  p^{\text{in}} \\
                  x^{\text{in}} \\
                \end{array}
              \right)+\frac{\kappa^{2}}{\sqrt{3}} \left(
                                      \begin{array}{c}
                                        -q_{\text{s},f_{\text{q}}}^{\text{in}} \\
                                         q_{\text{c},f_{\text{q}}}^{\text{in}} \\
                                      \end{array}
                                    \right)
 .
\end{eqnarray}\\
\\Eq.~(\ref{Eq:Hamiltonian}) shows that the interaction between the atomic ensemble and the light field leads to entanglement between these two systems. The exact form of the entangled modes and the degree of entanglement depends on the specific parameters characterizing the interaction, $\gamma_s$ and $Z$, and on the added noise. In essence, the underlying physics can be understood by considering the special case of an Einstein-Podolski-Rosen-entangled state~\cite{EPR} and the nonlocal variables $x_{-}=\left(x_{\text{\tiny{atom}}}- x_{\text{\tiny{light}}}\right)/\sqrt{2}$ and $p_{+}=\left(p_{\text{\tiny{atom}}}+ p_{\text{\tiny{light}}}\right)/\sqrt{2}$, where $x_{\text{\tiny{atom}}}$, $p_{\text{\tiny{atom}}}$ and $x_{\text{\tiny{light}}}$, $p_{\text{\tiny{light}}}$ are appropriate quadratures of the atomic ensemble and a suitable integrated light mode. The inequality  $\text{var}(x_{-})+\text{var}(p_{+})< 1$ indicates that the two systems are entangled. It certifies the existence of intersystem correlations which are stronger than classically allowed. For a perfectly entangled state $\text{var}(x_{+})+\text{var}(p_{-})=0$. In this case, the $x$-quadratures of the two systems are perfectly correlated. If a measurement of the atomic $x$-quadrature yields the value $x_1$, then the corresponding measurement on the photonic system yields also $x_1$. The $p$-quadratures are anti-correlated. If the value $p_1$ is obtained in a measurement of the atomic ensemble, the corresponding measurement on the light field yield $-p_1$.
\subsection{Teleportation scheme}\label{Sec:SI:TeleportationScheme}
In this section, we discuss the teleportation scheme. In Sec.~\ref{Sec:SI:TeleprtationProtocol&IORs}, we explain the basic working principle of the protocol and provide the corresponding input-output relations. In Sec.\ref{Sec:SI:Fidelity}, we compute the teleportation fidelity.
\subsubsection{Protocol and input-output relations}\label{Sec:SI:TeleprtationProtocol&IORs}
A standard teleportation scheme involving the three parties Alice,
Bob and Charlie consists of the following three steps, which allow
Alice to teleport a quantum state provided by Charlie to Bob. (i)
Alice and Bob establish an entangled link, which is shared between
the two remote parties. (ii) Alice performs a Bell measurement on
her part of the entangled state shared with Bob and an unknown
quantum state prepared by Charlie. (iii) Alice uses a classical
channel to communicate the measurement outcome to Bob, who
performs a local operation on his quantum state conditioned on
Alice's result.\\
\\The setup used here is shown in Fig.1a in the main text. The quantum
state prepared by Charlie (on Alice's side) is stored in ensemble A. This state is teleported to ensemble B, which
represents Bob, while the light field in $x$-polarization plays the role of Alice. Step (i)
in the standard protocol outlined above corresponds to the
interaction between the light field and the first atomic ensemble
which results in an entangled state. The distribution of
entanglement between the two remote sites is realized by means of
the free propagation of the photonic state. Step (ii) corresponds
to the interaction of the light field with the second ensemble and
the subsequent measurement of the $y$-quadrature of the
transmitted light by means of homodyne detection. Step (iii) is
implemented in the form of a feedback operation realizing a
conditional displacement on ensemble B using radio-frequency
magnetic fields.
Note that the measurement result in step (ii) is probabilistic and leads therefore to a random displacement of Bob's state in phase space. Since the measurement result is known, the resulting $x$, and $p$-quadratures of ensemble B are known such that a conditional displacement operation can be applied in step (iii) which shifts Bob's state to the desired coordinates in phase space. In principle, it is not necessary to perform the displacement operation on ensemble B. Instead, the outcome of the measurement in step (ii) can also be communicated to Bob who uses this information to locate the quantum state correctly in phase space when reading it out.\\
\\The measured values of $y_{\text{c},-} ^{\text{out}}$ and $y_{\text{s},-} ^{\text{out}}$ are fed back onto ensemble B as explained above. Due to symmetry reasons, applying equal gain factors $g_x=g_p=g$ is optimal. This yields
\begin{eqnarray}\label{Eq:SI:TeleportationNoisyIORs}
\left(
  \begin{array}{c}
    x_{\text{B}}^{\text{tele}} \\
    p_{\text{B}}^{\text{tele}} \\
  \end{array}
\right)&=&\bar{c}_{\text{B}}\left(
  \begin{array}{c}
    x_{\text{B}}^{\text{in}} \\
    p_{\text{B}}^{\text{in}} \\
  \end{array}
\right)+\bar{c}_{\text{A}}\left(
  \begin{array}{c}
    x_{\text{A}}^{\text{in}} \\
    p_{\text{A}}^{\text{in}} \\
  \end{array}
\right)+\bar{c}_{\text{N,B}}\left(
                             \begin{array}{c}
                               F^{\text{in}}_{x\text{B}} \\
                               F^{\text{in}}_{p\text{B}} \\
                             \end{array}
                           \right)
+\bar{c}_{\text{N,A}}\left(
                             \begin{array}{c}
                               F^{\text{in}}_{x\text{A}} \\
                               F^{\text{in}}_{p\text{A}} \\
                             \end{array}
                           \right)\nonumber\\
&+&
\bar{c}_{\text{y}}\left(
  \begin{array}{c}
    y_{\text{s},\bar{f}_{\text{y}}}^{\text{in}} \\
    y_{\text{c},\bar{f}_{\text{y}}}^{\text{in}} \\
  \end{array}
\right)+\bar{c}_{\text{q}}\left(
  \begin{array}{c}
    q_{\text{c},\bar{f}_{\text{q}}}^{\text{in}} \\
    -q_{\text{s},\bar{f}_{\text{q}}}^{\text{in}} \\
  \end{array}
\right).
\end{eqnarray}
The modes appearing in this equation are defined as in Eq.~(\ref{Eq:SI:GeneralModeDefinitions}) with
\begin{eqnarray*}
\bar{f}_{\text{N,B}}&=&\sqrt{2\gamma_{\text{extra}}}(e^{-\gamma(T-t)}+\frac{Z\sqrt{2\gamma_{\text{s}}\gamma} g}{\sqrt{1-e^{-2\gamma T}}}\int_t^Tdt' e^{-\gamma(2t'-t)}[1-2\gamma_{\text{s}}(t'-t)]),\\
\bar{f}_{\text{N,A}}&=&\frac{Z\sqrt{\gamma_{\text{s}}\gamma_{\text{extra}}}}{\sqrt{\gamma}}\frac{g}{\sqrt{1-e^{-2\gamma T}}}\left(e^{-\gamma t}-e^{-2\gamma T}e^{\gamma t}\right),\\
\bar{f}_{\text{y}}&=&-\frac{\sqrt{2\gamma_{\text{s}}}}{Z}e^{-\gamma(T-t)}+\frac{\gamma_{\text{s}}}{\sqrt{\gamma}}\frac{g}{\sqrt{1-e^{-2\gamma T}}}\{\frac{2\gamma}{\gamma_{\text{s}}}e^{-\gamma t}-[e^{-\gamma t}(2-\frac{\gamma_{\text{s}}}{\gamma})\\
&-&e^{-2\gamma T} e^{\gamma t}(2-\frac{\gamma_{\text{s}}}{\gamma}-2\gamma_{\text{s}}(T-t))]\},\\
\bar{f}_{\text{q}}&=&\sqrt{2\gamma_{\text{s}}}Z\{e^{-\gamma(T-t)}+\frac{Zg\sqrt{\gamma_\text{s}}}{\sqrt{2\gamma}\sqrt{1-e^{-2\gamma T}}}[e^{-\gamma t}(2-\frac{\gamma_{\text{s}}}{\gamma})\\
&-&e^{-2\gamma T} e^{\gamma t}(2-\frac{\gamma_{\text{s}}}{\gamma}-2\gamma_{\text{s}}(T-t))]\}.\\
\end{eqnarray*}
The corresponding normalization factors read
\begin{eqnarray*}
\bar{N}_{\text{N,B/A}}&=&\int_0^Tdt\bar{f}_\text{N,B/A}(t)^2,\ \ \ \ \
\bar{N}_{\text{y/q}}=\frac{1}{2}\int_0^Tdt\bar{f}_\text{y/q}(t)^2.
\end{eqnarray*}
and the coefficients appearing in Eq.~(\ref{Eq:SI:TeleportationNoisyIORs}) are given by
\begin{eqnarray*}
\bar{c}_{\text{B}}&=&e^{-\gamma T}+\frac{gZ\sqrt{\gamma_{\text{s}}}}{\sqrt{2\gamma}}\left(\sqrt{1-e^{-2\gamma T}}[1-\frac{\gamma_{\text{s}}}{\gamma}]+2\gamma_{\text{s}}T\frac{e^{-2\gamma T}}{\sqrt{1-e^{-2\gamma T}}}\right),\\
\bar{c}_{\text{A}}&=&\frac{gZ\sqrt{\gamma_{\text{s}}}}{\sqrt{2\gamma}}\sqrt{1-e^{-2\gamma T}},\\
\bar{c}_{\text{N,B}}&=&\sqrt{\bar{N}_{\text{N,B}}},\ \ \
\bar{c}_{\text{N,A}}=\sqrt{\bar{N}_{\text{N,A}}},\ \ \
\bar{c}_{\text{y}}=\sqrt{\bar{N}_{\text{y}}},\ \ \
\bar{c}_{\text{q}}=\sqrt{\bar{N}_{\text{q}}}.
\end{eqnarray*}
All coefficients and mode functions in the input-output equation for the final atomic state Eq.~(\ref{Eq:SI:TeleportationNoisyIORs}) carry a bar in order to avoid confusion with the coefficients and mode functions appearing in the readout equation Eq.~(\ref{Eq:SI:Read-out}) in Sec.~\ref{Sec:SI:Interaction}.
\subsubsection{Teleportation fidelity}\label{Sec:SI:Fidelity}
The performance of the protocol is assessed using the average
fidelity with respect to a Gaussian distribution of coherent input
states as figure of merit. The fidelity $F=|\langle
\Psi_{\text{B}}^{\text{tele}}|\Psi_\text{B}^{\text{opt}}\rangle|^2$ is given by the overlap of
the final atomic state in ensemble B (Bob's state), $|\Psi_{\text{B}}^{\text{tele}}\rangle$ , which is described by
$x_{\text{B}}^{\text{tele}}$ and $p_{\text{B}}^{\text{tele}}$ and the optimal final state which is
defined by the initial state in ensemble A (Charlie's state) $x_{\text{A}}^{\text{in}}$, $p_{\text{A}}^{\text{in}}$. For a
given coherent input state with mean values $\langle x_{\text{A}}\rangle$
and $\langle p_{\text{A}}\rangle$ (and variances
$\text{var}(x_{\text{A}})=\text{var}(p_{\text{A}})=1/2$ ), the single-shot fidelity
is given by
\begin{eqnarray*}
F(\langle x_{\text{A}}\rangle,\langle p_{\text{A}}\rangle)&=&2\
\frac{e^{\frac{-\left(|\langle
x_{\text{A}}\rangle|-|\langle
x_{\text{B}}^{\text{tele}}\rangle|\right)^2}{1+2\text{var}(
x_{\text{B}}^{\text{tele}})}}e^{\frac{-\left(|\langle
p_{\text{A}}\rangle|-|\langle
p_{\text{B}}^{\text{tele}}\rangle|\right)^2}{1+2\text{var}
(p_{\text{B}}^{\text{tele}})}}} {\sqrt{\left(1+2\text{var}(
x_{\text{B}}^{\text{tele}})\right)\left(1+2\text{var}( p_{\text{B}}^{\text{tele}})\right)}}\ ,
\end{eqnarray*}
such that the average fidelity $\bar{F}(\bar{n})$ with respect to a
Gaussian distribution with width $\bar{n}$,
\begin{eqnarray}\label{Eq:Fidelity}
\bar{F}(\bar{n})\!\!&=&\!\!\frac{1}{2\pi \bar{n}}\!\int_{-\infty}^{\infty}
\!\!\!\!\!\!\!\!\!\!\!\!\int d\langle x_{\text{A}}\rangle d\langle
p_{\text{A}}\rangle F\!\left(\!\langle x_\text{A}\rangle,\!\langle
p_\text{A}\rangle\!\right)\! e^{-\frac{\langle x_{\text{A}}\rangle^2+\langle
p_{\text{A}}\rangle^2}{2\bar{n}}},\\
\!\!\!\!&=&\!\!\!\!\frac{\sqrt{2}}{\sqrt{1\!+\!2\text{var}\left(
x_{\text{B}}^{\text{tele}}\right)\!+\!2\bar{n}\!\left(1\!-\!\left|\frac{\langle x_{\text{B}}\rangle}{\langle x_{\text{A}}\rangle}\right|\right)^2\!}}\frac{\sqrt{2}}{\sqrt{1\!+\!2\text{var}\left(
p_{\text{B}}^{\text{tele}}\right)\!+\!2\bar{n}\!\left(1\!- \!\left|\frac{\langle p_{\text{B}}\rangle}{\langle p_{\text{A}}\rangle}\right|\right)^2}}.\nonumber
\end{eqnarray}
Both ensembles are initialized in a coherent spin state with $\langle x_{\text{B}}\rangle=\langle x_{\text{A}}\rangle=0$ and $\langle x_{\text{B}}^2\rangle=\langle x_{\text{A}}^2\rangle=1/2$. The photonic modes are also initially in the vacuum state, such that
\begin{eqnarray*}
\left|{\langle x_{\text{B}}\rangle}/{\langle x_{\text{A}}\rangle}\right|&=&
\left|{\langle p_{\text{B}}\rangle}/{\langle p_{\text{A}}\rangle}\right|=\bar{c}_{\text{A}},\\
\text{var}(x_{\text{B}}^{\text{tele}})&=&\text{var}(p_{\text{B}}^{\text{tele}})=\frac{1}{2}\left(\bar{c}_{\text{B}}^2+\bar{c}_{\text{A}}^2
+m\ \!\bar{c}_{\text{N,B}}^2+m\ \!\bar{c}_{\text{N,A}}^2+\bar{c}_{\text{y}}^2+\bar{c}_{\text{q}}^2\right).
\end{eqnarray*}
Fig.~(S1) shows the average teleportation fidelity $\bar{F}(n)$ in comparison to the classical limit $F_{\text{clas}}=\left(1+\bar{n}\right)/\left(1+2\bar{n}\right)$, which
cannot be surpassed by classical means~\cite{FuSBFKP98,BrFK00,HaWPC05}.
This figure also displays the average fidelity which can be achieved using a QND-interaction (i.e. for very large detuning, see Sec.~\ref{Sec:SI:Interaction}) if the classical driving pulses are modulated in time. These results have been obtained by considering different exponential functions $f_{\text{B}}(t)\propto e^{f_{\text{\tiny{B}}}t}$ and $f_{\text{A}}(t)\propto e^{f_{\text{\tiny{A}}}t}$ for the pulse shape of the classical field in the first and second interaction. The fidelity is optimized with respect to $f_{\text{\tiny{B}}}$ and
$f_{\text{\tiny{A}}}$. Fig.~(S1) b shows that fidelities close to one can be obtained in principle.

\begin{figure}\label{Fig:SI:FidelityIdeal}
\begin{center}
\includegraphics[width=0.8\columnwidth]{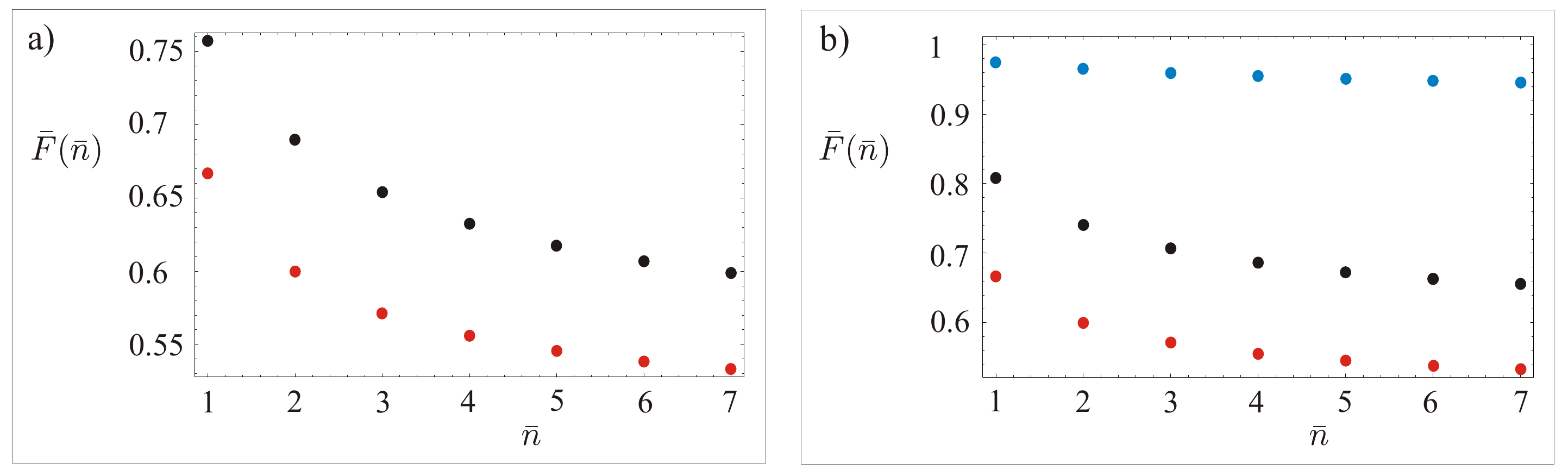}
 \caption{Average teleportation fidelity $\bar{F}(\bar{n})$ for optimal gain $g$ versus width of the distribution of input states $\bar{n}$. The lowest (red) line in both panels represents the classical benchmark. a) Teleportation fidelity for the measured experimental parameters (see methods). b) Maximum teleportation fidelity in the absence of losses. The curve in the middle (black) shows the attainable teleportation fidelity for the interaction used in the experiment ($Z=2.5$). The upmost curve (blue) depicts the QND-fidelity (for $Z\rightarrow\infty$, $\gamma_{\text{s}}TZ^2=\text{const}$) for exponentially shaped driving pulses (compare text, Sec.~\ref{Sec:SI:Fidelity}). }
 \end{center}
 \end{figure}


\begin{thebibliography}{54}

\bibitem{bri98} Briegel, H.-J., D\"ur, W., Cirac, J. I. \& Zoller, P. Quantum Repeaters: The Role of Imperfect Local Operations in Quantum Communication. {\it Physical Review Letters}, {\bf 81}, 5932--5935 (1998).
\bibitem{riedmatten04} de Riedmatten, H. {\it et al.} Long Distance Quantum Teleportation in a Quantum Relay Configuration {\it Physical Review Letters}, {\bf 92}, 47904 (2004).
\bibitem{gottesman99}Gottesman, D. \& Chuang, I.L.  Demonstrating the viability of universal quantum computation using
    teleportation and single-qubit operations {\it Nature}, {\bf 402}, 390--393 (1999).
\bibitem{gottesman01} Gottesman, D., Kitaev, A., Preskill, J. Encoding a qubit in an oscillator. {\it Physical Review A}, {\bf 64}, 012310  (2001).
\bibitem{bennet93} Bennett, C. H. {\it et al.} Teleporting an unknown quantum state via dual classical and Einstein-Podolsky-Rosen channels {\it Physical Review Letters}, {\bf 70 }, 1895--1899 (1993).
\bibitem{vaidman94}  Vaidman, L. Teleportation of quantum states {\it Physical Review A}, {\bf 49}, 1473--1476  (1994).
\bibitem{bouw1997} Bouwmeester, D. {\it et al.}, A Experimental quantum teleportation {\it Nature}, {\bf 390}, 575--579 (1997).
\bibitem{Furusawa98} Furusawa, A. {\it et al.} Unconditional Quantum Teleportation {\it Science }, {\bf 282}, 706--709 (1998).
\bibitem{riebe04} Riebe, M. {\it et al.}  Deterministic quantum teleportation of atomic qubits {\it Nature}, {\bf 429}, 737--739 (2004). Barret, M.D. {\it et al.}  Deterministic quantum teleportation with atoms {\it Nature}, {\bf 429}, 734--737 (2004).
\bibitem{she06} Sherson, J. {\it et al.} Quantum teleportation between light and matter {\it Nature}, {\bf 443}, 557--560 (2006).
\bibitem{Chen2008} Chen, Y.A. {\it et al.}  Memory-built-in quantum teleportation with photonic and atomic qubits {\it Nature Physics}, {\bf 4}, 103--107 (2008).

\bibitem{Olmschenk2009} Olmschenk, S. {\it et al.}  Quantum Teleportation Between Distant Matter Qubits {\it Science}, {\bf 323 }, 486--489 (2009).
\bibitem{Rempe2012} N\"olleke, C., Neuzner, A., Reiserer, A., Hahn, C., Rempe, G., \& Ritter, S. Efficient teleportation between remote single-atom quantum memories. Preprint available at http://arxiv.org/abs/1212.3127.
\bibitem{pan12} Bao X-H. {\it et al.} Quantum teleportation between remote atomic-ensemble quantum memories {\it Proceedings of the National Academy of Sciences of the United States of America}, {\bf 109}, 20347--20351 (2012).
\bibitem{dua01} Duan, L.M., Lukin, M.D., Cirac, J.I. \& Zoller, P. Long-distance quantum communication with atomic ensembles and linear optics {\it Nature}, {\bf 414}, 413--418 (2001).
\bibitem{zeilinger12}  Ma, X.S. {\it et al.} Quantum teleportation over 143 kilometres using active feed-forward {\it Nature}, {\bf 489 }, 269--273 (2012).
\bibitem{furusawa11}  Lee, N. {\it et al.}  Teleportation of Nonclassical Wave Packets of Light {\it Science}, {\bf 332}, 330--333 (2011).
\bibitem{ham08} Hammerer, K., S{\o}rensen, A.S. \& Polzik, E.S. Quantum interface between light and atomic ensembles {\it Reviews of Modern Physics}, {\bf 82}, 1041--1093 (2010).
\bibitem{was09} Wasilewski, W. {\it et al.} Generation of two-mode squeezed and entangled light in a single temporal and spatial mode {\it Optics Express}, {\bf 17}, 14444--14457 (2009).
\bibitem{kra10} Krauter, H. {\it et al.} Entanglement Generated by Dissipation and Steady State Entanglement of Two Macroscopic Objects {\it Physical Review Letters}, {\bf 107}, 080503 (2011).
\bibitem{she07} Sherson, J., Julsgaard, B. \& Polzik, E.S. Deterministic Atom-Light Quantum Interface {\it Advances in Atomic, Molecular, and Optical Physics}, {\bf 54}, 81--130 (2006).
\bibitem{ham05} Hammerer, K., Wolf, M.M., Polzik, E.S. \& Cirac, J.I. Quantum benchmark for storage and transmission of coherent states {\it Physical Review Letters}, {\bf  94}, 150503 (2005).
\bibitem{sm} More details are given in the Supplementary information.
\bibitem{muschik13} Muschik, C.A., Hammerer, K., Polzik, E.S. \& Cirac J.I. Quantum Teleportation of Dynamics and Effective Interactions Between Remote Systems arXiv:1304.0319 (2013).
\bibitem{muschik2012} Muschik, C.A. {\it et al.} Robust entanglement generation by reservoir engineering {\it Journal of Physics B: Atomic, Molecular and Optical Physics}, {\bf 45}, 124021 (2012).
\bibitem{Vasilyev12} Vasilyev, D.V., Hammerer, K., Korolev, N. \& S{\o}rensen, A.S. Quantum Noise for Faraday Light Matter Interfaces {\it Journal of Physics B: Atomic, Molecular and Optical Physics} {\bf 45} 124007 (2012).
\end{thebibliography}

\begin{thebibliography}{54}
%
\bibitem{Silberfarb2003}
A. Silberfarb, I.H. Deutsch, Phys. Rev. A \textbf{68}, 13817 (2003).
%
\bibitem{Madsen2004}
L.B. Madsen, K. M\o{}lmer, Phys. Rev. A, \textbf{70}, 052324 (2004).
%
\bibitem{HPC05}
K. Hammerer, E.S. Polzik, and J.I. Cirac, Phys. Rev. A \textbf{72}, 052313 (2005).
%
\bibitem{MKJPCP12}
C.A. Muschik, H. Krauter, K. Jensen, J.M. Petersen, J.I. Cirac, and E.S. Polzik, J. Phys. B: At. Mol. Opt. Phys. \textbf{45}, 124021 (2012).
%
\bibitem{HSP10}
K. Hammerer, A. S\o{}rensen, and E.S. Polzik, Rev. Mod. Phys.
\textbf{82}, 1041 (2010).
%
\bibitem{MPC11}
C.A. Muschik, E.S. Polzik, and J.I. Cirac, Phys. Rev. A \textbf{83}, 052312 (2011).
%
\bibitem{NoisePaper}
D. V. Vasilyev, K. Hammerer, N. Korolev, A. S. Sorensen, J. Phys. B: At. Mol. Opt. Phys. \textbf{45}, 124007 (2012).
%
\bibitem{Denis}
D. V. Vasilyev and Klemens Hammerer, personal communication.
%
\bibitem{EPR}
A. Einstein, B. Podolsky and N. Rosen, Phys. Rev. \textbf{47}, 777 (1935).
%
\bibitem{FuSBFKP98}
A. Furusawa, J.L. S\o{}rensen, S. L. Braunstein, C. A. Fuchs, H.J. Kimble and E.S. Polzik, Science \textbf{282}, 706 (1998).
%
\bibitem{BrFK00}
S.L. Braunstein, H.J. Kimble and C.A. Fuchs, J. Mod. Opt. \textbf{47},267 (2000).
%
\bibitem{HaWPC05}
K. Hammerer, M. M. Wolf, E. S. Polzik and J. I. Cirac, Phys. Rev. Lett. \textbf{94}, 150503  (2005).
%
\end{thebibliography}
\end{document}